\documentclass[12pt]{article}

\usepackage{epsfig,latexsym,amsfonts,amsmath,amsthm,amssymb, mathrsfs,amsbsy,multirow,slashed,wasysym,textcomp,subfigure,hyperref,wrapfig,datetime,comment,mathtools,cancel,cite}
\usepackage[font={footnotesize,sl},bf]{caption}
\usepackage{comment}

\def\be{\begin{equation}}
\def\ee{\end{equation}}
\def\bea{\begin{eqnarray}}
\def\eea{\end{eqnarray}}
 \newcommand{\badat}{\begin{alignedat}}
 \newcommand{\eadat}{\end{alignedat}}

\long\def\new#1\endnew{{\bf #1}}		
\long\def\del#1\enddel{}

\def\del{\partial}

\usepackage{xcolor}
\definecolor{oldmauve}{rgb}{0.4, 0.19, 0.28}
\definecolor{pansypurple}{rgb}{0.47, 0.09, 0.29}
\definecolor{burgundy}{rgb}{0.5, 0.0, 0.13}
\definecolor{carminepink}{rgb}{0.92, 0.3, 0.26}
\definecolor{blue(pigment)}{rgb}{0.2, 0.2, 0.6}
\definecolor{darkseagreen}{rgb}{0.56, 0.74, 0.56}
\definecolor{darkspringgreen}{rgb}{0.09, 0.45, 0.27}
\definecolor{ceruleanblue}{rgb}{0.16, 0.32, 0.75}

\begin{document}
%\title{}
%\date{}
\numberwithin{equation}{section} % equation numbers follow sections

\begin{titlepage}
  \thispagestyle{empty}
  
  \begin{center} 
  \vspace*{2cm}
{\LARGE{The disk 1-point function in timelike Liouville theory}}

\vskip1cm

\centerline{Gaston Giribet$^1$ and Bruno Sivilotti$^{2,3}$}
   
\vskip1cm

{${}^1$ Center for Cosmology and Particle Physics\\ Department of Physics, New York University,\\ {\it 726 Broadway, New York City, NY10003, USA.}}\\
\vskip0.5cm
{${}^2$ Universidad de Buenos Aires\\ Departamento de F\'isica and IFIBA - CONICET\\
1428 Buenos Aires, Argentina}\\
\vskip0.5cm
{${}^3$ Perimeter Institute for Theoretical Physics\\ {\it 31 Caroline Street North, Waterloo, ON N2L2Y5, Canada.}}\\
\vskip0.5cm{E-mail: \texttt{gaston.giribet@nyu.edu}, \texttt{bsivilotti@perimeterinstitute.ca}}

\end{center}

\vskip1cm

\begin{abstract}
We compute the disk 1-point function in timelike Liouville theory. Using the Coulomb gas formalism and analytically continuing in the number of screening operators, we derive an explicit formula, which is shown to satisfy the correct reflection symmetry, to have the expected self-dual properties, to fulfill the bootstrap shift-equations, and to reduce to previous known results in the appropriate limits. In the limit of zero cosmological constant, our result reproduces the one recently obtained in \cite{Santachiara}.
\end{abstract}

\end{titlepage}

\section{Introduction}

Timelike Liouville theory is an interesting 2-dimensional quantum field theory whose properties we are only just beginning to understand in detail. It is usually referred to as imaginary Liouville theory or imaginary Gaussian multiplicative chaos. It is related to many topics in physics, such as string theory on time-dependent backgrounds and quantum gravity. Among the most interesting applications of this theory, there is its application to cosmology in de Sitter (dS) space, which has recently received attention. For example, in \cite{Anninos} timelike Liouville field theory in the disk geometry was studied as a tractable toy model for quantum cosmology. The authors considered a model in which a unitary 2-dimensional conformal field theory is coupled to a fluctuating metric governed by a gravitational action with a positive cosmological constant. After fixing the conformal gauge for the metric and integrating out the matter degrees of freedom, the action takes the form of timelike Liouville field theory \cite{HMW}, which corresponds to the theory with central charge $c_L<1$, cf. \cite{RibaultSantachiara}. They observed that the distinctive feature of the timelike theory, namely the fact that the kinetic term has the wrong sign, is reminiscent of the gravitational action being unbounded from below, which reinforces the analogy with quantum gravity. In this setup, the 1-point function of an exponential primary operator inserted in the disk geometry with conformally invariant boundary conditions turns out to be related to solutions of the Wheeler-de Witt equation. By choosing the complexified integration contour judiciously, the path integral calculation of the disk observable is reminiscent of the Hartle-Hawking no-boundary wave function. 

A puzzling feature of this calculation is that the semiclassical result proposed in \cite{Anninos} seems to be in conflict with previous bootstrap calculations of the timelike 1-point functions. In \cite{Bautista}, a specific linear combination of the bootstrap shift equation, which differs from the result proposed in \cite{Anninos}, was argued to give the right disk 1-point function in timelike Liouville field theory. According to the quantum cosmology interpretation, the expression obtained in \cite{Bautista} would rather correspond to a specific linear combination of the Hartle-Hawking and the Vilenkin wave functions, being different from the no-boundary proposal. Such a solution for the timelike theory is not the naive analytic continuation of its spacelike counterpart.

Here, aiming at contributing to this discussion, we give an alternative computation of the disk 1-point function in timelike Liouville theory. Resorting to the Coulomb gas approach and performing an astute analytic continuation of the integral expressions obtained, we obtain an expression for this observable which passes a series of non-trivial consistency checks: it satisfies the correct reflection symmetry, has the expected self-dual properties, satisfies the bootstrap shift-equations, and reduces to previous known results in the appropriate limits. The justification to perform the computation resorting to the Coulomb gas formalism comes from the fact that this method has already proven to work well in the computation of the 3-point correlation functions in the timelike theory on the sphere \cite{Giribet}. Performing the analogous calculation for the disk 1-point function will allow us to investigate further a peculiar feature that, as observed in \cite{Giribet}, appears in the integration over the zero-mode of the timelike field. Besides, the Coulomb gas approach makes contact with the path integral formulation of the field theory.

The disk 1-point function in spacelike Liouville field theory was computed in \cite{FZZ}. Technical aspects of the calculation were further discussed in \cite{Teschner}. More recently, this quantity was studied in a rigorous approach \cite{ChatterjeeWitten, Vargas} in references \cite{Remy, Remy2, Remy3}; see also \cite{Huang:2018ncv}. In the case of the timelike theory \cite{HMW}, the disk 1-point function was originally studied in \cite{Strominger}; more recently, a calculation of this quantity was given in \cite{Bautista} using bootstrap methods. In a more formal approach, timelike correlation functions were recently studied in \cite{Santachiara, Chatterjee, Chatterjee:2026zmb}. Timelike Liouville theory and its related Complex Liouville theory have recently appeared in the physics literature in relation to quantum gravity in 2 and 3 dimensions, among many other topics \cite{Bautista:2015wqy, Bautista:2019jau, Collier:2023cyw, Collier:2024kmo, Collier:2024kwt, Collier:2024lys, Collier:2024mlg, Verlinde:2024zrh, Blommaert:2025eps, Bruno, Giribet:2022cvw, Anninos:2021ene}; for more applications of the timelike theory see \cite{Hashimoto:2022dro, Giribet:2023gub, Balthazar:2023oln, Chu:2026rle} and references thereof.

This papers is organized as follows: In section 2, we present the detailed derivation of the disk 1-point function in spacelike Liouville theory. We solve the multiple Selberg-type integrals involved in this computation and discuss the steps leading to the analytic continuation and the final result. Although the calculation of the 1-point function in the Coulomb gas formalism is outlined in many places, the details are usually omitted. We carry out the full calculation and present the details to lay the groundwork that would later allow us to extend the method to the timelike case. Our calculation is an adaptation to Liouville theory of the one performed in reference \cite{Ribault} for Wess-Zumino-Witten (WZW) theory. In Section 3, we show that the disk 1-point function in the timelike Liouville theory can also be computed using the Coulomb gas formalism. We obtain an explicit and finite result in $1/c_L$. In section 4, we perform a series of consistency checks of our result. In addition, we discuss the different prescriptions to integrate the zero-mode. In particular, we discuss the prescription proposed in \cite{Santachiara} and how it leads to solving a problem observed in \cite{Giribet}. In section 5, we summarize our findings.

\section{Spacelike disk 1-point function}

\begin{figure}[h]
    \begin{center}
        \includegraphics[width=14
        cm]{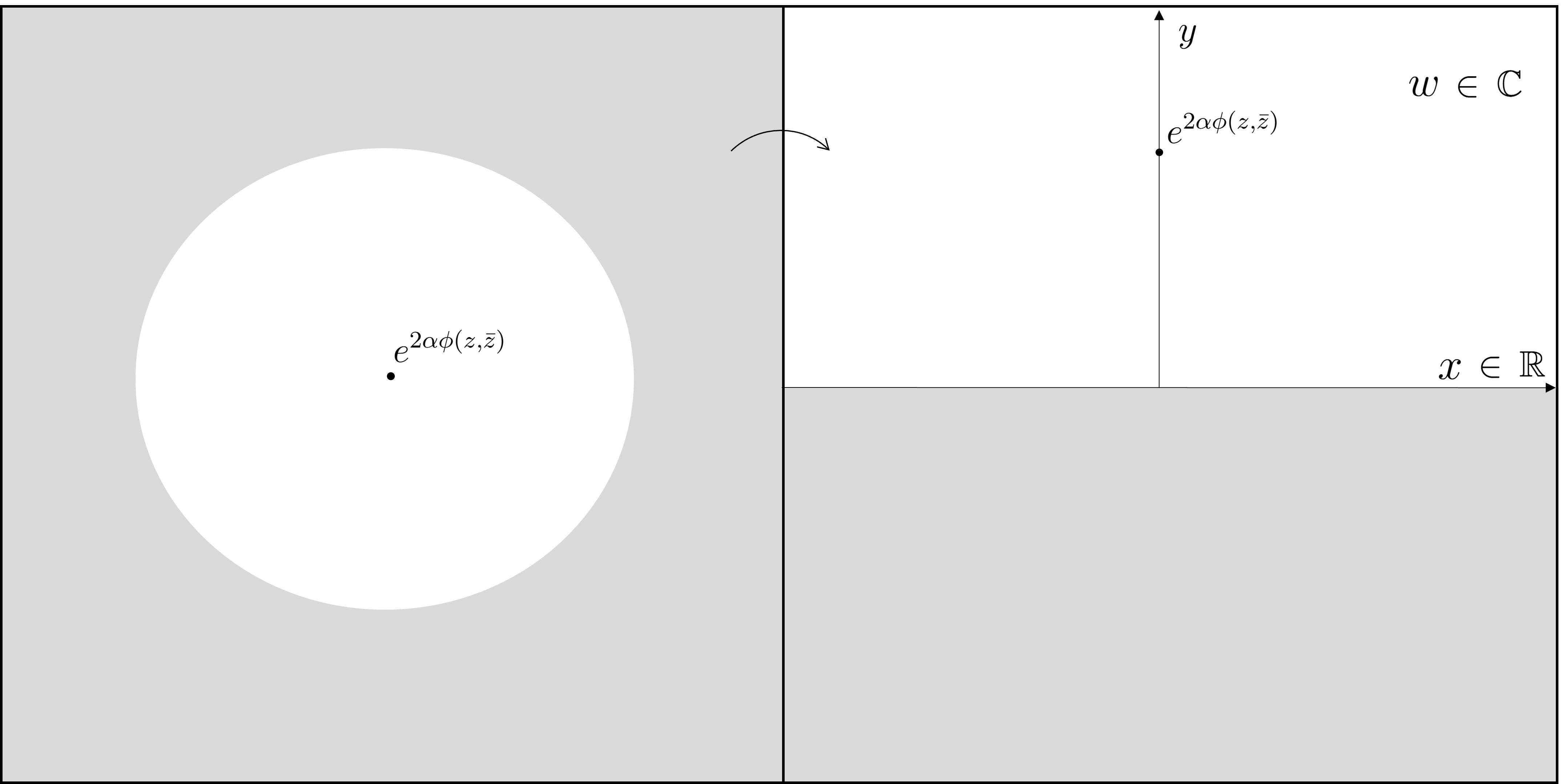}
        \caption{Scheme of the Liouville 1-point function.}
        \label{Figure1}
    \end{center}
\end{figure}

The action of the spacelike Liouville field theory is
\begin{equation}
S [\phi ; \Lambda , \Lambda_B]=\frac{1}{4\pi }\int_{\Sigma }d^2x\,\sqrt{g}\,\Big(g^{\mu\nu}\partial_{\mu}\phi\partial_{\nu}\phi+QR\phi+4\pi\Lambda\, e^{2b{\phi }}\Big)+B
\end{equation}
with the boundary term
\begin{equation}
B=\frac{1}{2\pi }\int_{\partial\Sigma}d\text{x}\,\sqrt{h}\,\Big(QK\phi+2\pi\Lambda_B\,e^{b\phi}\Big)
\end{equation}
and the relation
\begin{equation}
Q=b+\frac{1}{b}.
\end{equation}
In the expressions above, $R$ is the scalar curvature of the 2-dimensional manifold $\Sigma $ provided with a metric $g$; $h$ is the induced metric in the boundary $\partial \Sigma $, and $K$ is its extrinsic curvature. $\Lambda$ and $\Lambda_B$ are the so-called cosmological constant and boundary cosmological constant, respectively. This defines a unitary, non-rational conformal field theory of central charge
\begin{equation}
c_L=1+6Q^2\geq 25
\end{equation}
In this theory, we will compute the 1-point function of the primary operator 
\begin{equation}
V_{\alpha}(z,\bar z)\, :=\,  e^{2\alpha \phi (z,\bar z )}\label{vertex}
\end{equation}
on the disk geometry with conformally invariant boundary conditions; namely,
\begin{equation}
\Big\langle \,V_{\alpha}(z,\bar z)\,\Big\rangle\, :=\, \int\mathcal{D}\phi \, e^{-S[\phi ;\Lambda, \Lambda_B]}\, e^{2\alpha \phi (z,\bar z )}\label{function}
\end{equation}
The conformal dimension of the field created by the operator (\ref{vertex}) is $\Delta_{\alpha } = \alpha (Q-\alpha )$. This implies that the 1-point function will take the following form\footnote{Here, as in most of this work, we will assume to be working in the upper complex plane. This can always be done since one can conformally map the disk to the upper complex plane, as shown in Fig. \ref{Figure1}.}
\begin{equation}
\Big\langle \,V_{\alpha}(z, \bar z )\,\Big\rangle = \frac{\text{U}_{b}(\alpha ) }{{\ \ \big|z-\bar z \big|^{2\Delta_{\alpha}}}}\label{U}
\end{equation}
for some function $\text{U}_b(\alpha )$ that depends on $\alpha $ and also on $b$,  $\Lambda$, and $ \Lambda_B$. The Knizhnik–Zamolodchikov-Polyakov (KPZ) scaling of the correlator can be obtained by considering the shift in the field $\phi \to \phi + c$, which can be absorbed by rescaling $\delta \Lambda \simeq 2bc \Lambda$, $\delta \Lambda_B\simeq bc\Lambda_B$. This means that the 1-point function (\ref{U}) on the disk turns out to be proportional to $\langle e^{2\alpha \phi }\rangle \sim \Lambda^{(Q-2\alpha)/(2b)}$, and that $\Lambda \sim \Lambda^2_B$. That leaves the laborious task of finding out what the dependence is on $b$ and on the dimensionless ratio $\Lambda_B^2/\Lambda$.

The residues of this 1-point function can be computed using the Coulomb gas approach. Schematically, this takes the form \cite{FZZ}
\begin{equation}
\text{Res} \, \Big\langle \,V_{\alpha}(z,\bar z)\,\Big\rangle =\sum_{m=0}^{\infty } \sum_{l=0}^{\infty } \frac{(-\Lambda )^m}{m!}\frac{(-\Lambda_B )^l}{l!} \int\mathcal{D}\phi \, e^{-S[\phi;0,0]} e^{2\alpha \phi (z,\bar z )} \Big( \int d^2w\,e^{2b\phi (w,\bar w )}\Big)^m\Big( \int dx\,e^{b\phi (x)}\Big)^l \label{expression}
\end{equation}
with the restriction
\begin{equation}
\frac{Q-2\alpha}{2b}-m-\frac{l}{2}\, =\, 0\, . \label{charge}
\end{equation}
We will find convenient to define 
\begin{equation}
s\,:=\,\frac{Q-2\alpha}{2b}. \label{screening}
\end{equation}
Expression (\ref{expression}) follows from expanding the interaction terms in the action. Notice that the correlator on the right hand side of (\ref{expression}) is to be computed in the free theory. Condition (\ref{charge}) can be obtained by integrating the zero-mode of the Liouville field. This is the condition required for the marginal operators to screen the background charge. More precisely, the residues of the 1-point function when $2s \in \mathbb{Z}_{\geq 0}$ take the following form
\begin{eqnarray}\label{residues_specified}
\text{Res}\hspace{2pt}_{2s\in \mathbb{Z}_{\geq 0}} \, \Big\langle \,V_{\alpha}(z,\bar z)\,\Big\rangle\, &=&\, \sum_{m=0}^{\infty } \sum_{l=0}^{\infty } \frac{(-\Lambda )^m}{m!}\frac{(-\Lambda_B )^l}{l!} \, \prod_{i=1}^{m}\int d^2w_i\,
\prod_{k=1}^{l}\int dx_k\,\,\int\mathcal{D}\phi \, \Big( e^{-S[\phi;0,0]}\, e^{2\alpha \phi (z,\bar z )}\,  \nonumber \\ 
&&\ \ \ \  \prod_{i=1}^{m}e^{2b\phi(w_i,\bar w_i)}\prod_{k=1}^{l}e^{b\phi(x_k)} \, \Big)\,\times \, \delta _{s - m-\frac l2}
\end{eqnarray}
where the $\delta $-function makes it explicit that $Q/(2b)-\alpha /b = m+l/2$. As it is written, this expression makes sense only for $m\in \mathbb{Z}_{\geq 0}$ and $l\in \mathbb{Z}_{\geq 0}$, which implies the discrete values $\alpha =(1-2m-l)b/2+1/(2b)$. These values correspond to what is often called a resonant correlator. This expression precisely computes the residues of the simple poles of the 1-point function at these particular values of $\alpha$. However, the final result will take a form that allows us to analytically continue to complex values of $\alpha$; see (\ref{cuasifinal}) below. From now on, to simplify the notation, we will denote the quantity \eqref{residues_specified} as $\text{Res}\, \langle V_{\alpha}(z,\bar z)\rangle\,$ understanding that the residue is at $2s \in \mathbb{Z}_{\geq 0}$.

Without loss of generality, we can choose $z=iy$ with $y\in \mathbb{R}$. Using the free field correlator
\begin{equation}
\langle \phi(z,\bar z ) \phi(w,\bar w ) \rangle \, = \,-\log |z-w||\bar z - w| 
\end{equation}
we can explicitly compute the free field correlation function on the right hand side of the above expression. We obtain
\begin{eqnarray}
&&\text{Res} \, \Big\langle V_{\alpha}(iy,-iy)\Big\rangle = \sum_{m=0}^{\infty } \sum_{l=0}^{\infty } \frac{(-1)^{m+l}\Lambda ^m \,\Lambda_B^l}{|2y|^{2\alpha^2}\, m!\, l!} \prod_{i=1}^{m}\int d^2w_i
\prod_{k=1}^{l}\int dx_k \Big[ \Big( \prod_{k-1}^l (y^2+x_k^2)\prod_{i=1}^{m}|y^2+w_i^2|^2\Big)^{-2\alpha b}
\nonumber \\ 
&&\ \ \ \  \ \Big( \prod_{i=1}^{m} \prod_{k=1}^{l} |w_i-x_k|^2 \prod_{i'=1}^m\prod_{i=1}^{i'-1} |w_i-w_{i'}|^2 
\prod_{i'=1}^m\prod_{i=1}^{m} |w_i-\bar{w}_{i'}|
\prod_{k'=1}^l\prod_{k=1}^{k'-1} |x_k-x_{k'}|
\Big)^{-2b^2} \Big]\,\times \, \delta _{s - m-\frac l2}\nonumber
\end{eqnarray}
The second line corresponds to the Wick contractions of the screening operators among themselves, while the first line contains the contractions with the exponential primary field $e^{2\alpha \phi}$.  

The above integral can be explicitly solved. In order to show that, we first introduce the following integral formula \cite{Ribault}
\begin{eqnarray}
\mathcal{J}_{n,m}(a|y)&:=&\frac{1}{m!(n-2m)!}\int \prod_{i}^m d^2w_i\, |y^2+w_i^2|^{-2a}\int \prod_{k=1}^{n-2m}dx_k\, (y^2+x_k^2)^{-a}\, \Big( \prod_{i=1}^m \prod_{k=1}^{n-2m} |w_i-x_k|^2\nonumber \\
&& \ 
\prod_{i=1}^{m}\prod_{i'=1}^{i-1} |w_i-w_{i'}|^2
\prod_{i=1}^{m}\prod_{i'=1}^{m} |w_i-\bar{w}_{i'}|
\prod_{k=1}^{n-2m}\prod_{k'=1}^{k-1} |x_k-x_{k'}|\Big)^{-2b^2}=\nonumber \\
&& \ 
\frac{|2y|^{n(1-2a+(1-n)b^2)}}{n!}\Big(\frac{2\pi }{\Gamma(1-b^2)}\Big)^{n}\Big(\frac{1}{4\sin(\pi b^2)}\Big)^m\, I_n(a)\, J_{n,m}(a)
\end{eqnarray}
where
\begin{equation}
I_n(a):=\prod_{i=1}^{n-1}\frac{\Gamma(1-(i+1)b^2)\Gamma(2a-1+(n-1+i)b^2)}{\Gamma^2(a+ib^2)}
\end{equation}
and
\begin{equation}
J_{n,m}(a)=\sum_{i=0}^m\frac{(-1)^i (n-m-i)!}{(m-i)!(n-2m)!} \frac{\sin(\pi b^2(n+1-2i))}{\sin(\pi b^2(n+1-i))}\prod_{r=0}^{i-1}\frac{\sin(\pi b^2(n-r))\sin(\pi (a+(n-r)b^2))}{\sin(\pi b^2(r+1))\sin(\pi (a+rb^2))}
\end{equation}
We would need to evaluate these expressions for $a=2\alpha b$, $l=n-2m$, so that $n=2s=l+2m$. We have
\begin{eqnarray}
\text{Res} \, \Big\langle \,V_{\alpha}(iy, -iy)\,\Big\rangle &=&\sum_{m=0}^s \frac{(-1)^{2s-m}\Lambda^m\Lambda_B^{2s-2m}}{(2s)!}\, {|2y|^{2s(1-4\alpha b -(2s-1))b^2-2\alpha^2}} \nonumber \\
&&\Big(\frac{2\pi }{\Gamma(1-b^2)}\Big)^{2s}\Big(\frac{1}{4\sin (\pi b^2)}\Big)^{m}\, I_{2s}(2\alpha b)\, J_{2s,m}(2\alpha b)
\end{eqnarray}
where $2s(1-4\alpha b -(2s-1)b^2)=-2\alpha Q +4\alpha ^2$; recall, $s=(Q-2\alpha )/b$ and $Q=b+1/b$. Plugging this into the expressions above, we obtain
\begin{equation}
I_{2s}(2\alpha b )=\prod_{i=0}^{2s-1}\frac{\Gamma(1-(i+1)b^2)}{\Gamma(2\alpha b +ib^2)}
\end{equation}
which, using the relation among $s$, $m$ and $l$, after multiple cancellations of $\Gamma$-functions, simplifies to $I_{2s}(2\alpha b )=\Gamma(2\alpha b-b^2)$. Then, we arrive at the following expression
\begin{eqnarray}
\text{Res} \, \Big\langle \,V_{\alpha}(iy, -iy)\,\Big\rangle &=&|2y|^{-2\alpha (Q-\alpha )}\frac{\Gamma(2\alpha b-b^2)}{\Gamma(1+(Q-2\alpha)/b)}\Big(-\frac{2\pi \Lambda_B}{\Gamma(1-b^2)}\Big)^{\frac{Q-2\alpha }{b}} \nonumber \\
&&\ \sum_{m=0}^{s}\Big(-\frac{\Lambda}{\Lambda_B^2}\Big)^{m}\Big(\frac{1}{4\sin(\pi b^2)}\Big)^{m}\, J_{2s,m}(2\alpha b)
\end{eqnarray}
Now, we introduce a new parameter $\gamma$, which is implicitly  defined as follows
\begin{equation}
\cosh^2(\pi b \gamma )\, := \, \frac{\Lambda_B^2}{\Lambda }\sin(\pi b^2)
\end{equation}
This allows us to rewrite the expression above as
\begin{eqnarray}
\text{Res} \, \Big\langle \,V_{\alpha}(iy, -iy)\,\Big\rangle &=& |2y|^{2\alpha (\alpha -Q )}
\frac{\Gamma(2\alpha b-b^2)}{\Gamma(1+(Q-2\alpha)/b)}
\Big(\frac{\pi^2\Lambda}{\Gamma^2(1-b^2)\,\sin(\pi b^2)}\Big)^{\frac{Q-2\alpha }{2b}}\nonumber \\
&&\ \ \sum_{m=0}^{s}(-1)^{2s+m} \,(2\cosh (\pi b\gamma ))^{2s-2m}\,J_{2s,m}(2\alpha b)
\end{eqnarray}
In order to simplify this expression, we make use of the following formula\footnote{See Eq. (5.12) in \cite{Ribault}, where a similar calculation for the WZW model is performed. {Notice that our formula has a factor of $2$ that is not there in \cite{Ribault}. We checked the formula and believe the $2$ must be there.} }
\begin{eqnarray}
\sum_{m=0}^s(-1)^{m} (2\cosh(\pi b\gamma ))^{2s-2m} J_{2s,m}(2\alpha b)=2
\sum_{i=0}^{2s}\cosh(2\pi b\gamma(s-i))\, \times \nonumber
\\
 \ \ \ \ \ \ \ \ \ \ \ \ \ \ \ \ \ \ \ \ \ \ \prod_{r=0}^{i-1}\frac{\sin(\pi (2\alpha b -1+(r-1)b^2))\sin(\pi (rb^2-1))}{\sin(\pi b^2(r+1))\sin(\pi b (2\alpha + rb))}.
\end{eqnarray}
From this expression, we observe that only the summand with $i=0$ contributes; it is the only term that excludes the factor with $r=0$. This yields
\begin{equation}
\text{Res} \, \Big\langle \,V_{\alpha}(z, \bar z)\,\Big\rangle = {\big|z-\bar z\big|^{-2\Delta_{\alpha}}} \frac{2(-1)^{\frac{Q-2\alpha }{b}}\, \Gamma(2\alpha b-b^2)}{\Gamma(2+1/b^2-2\alpha/b)}\, \Big( \frac{\pi \Lambda \,\Gamma(b^2)}{\Gamma(1-b^2)}\Big)^{\frac{Q-2\alpha }{2b }}\cosh(\pi \gamma (Q-2\alpha))\label{cuasifinal}
\end{equation}
where we have used properties of the $\Gamma $-functions, such as
\begin{equation}\label{properties}
\Gamma(x)\Gamma(1-x)=\frac{\pi}{\sin(\pi x)}\, , \  \ \ \Gamma(x+1)=x\,\Gamma(x) \, , \ \ \ \Gamma(1+x-n)=(-1)^{n}\,\frac{\Gamma(1+x)\Gamma(-x)}{\Gamma(n-x)}
\end{equation}
with $x\in \mathbb{C}$, $n\in \mathbb{Z}_{\geq 0}$. These properties will be used in many steps of the calculation throughout the paper.

From \eqref{cuasifinal}, which gives the residues of the 1-point function on the disk for $2s\in \mathbb{Z}_{\geq 0}$, we can obtain the actual 1-point function by restoring the factor that accounts for the pole. To do this, we note that, using the third property of the $\Gamma$-function \eqref{properties} with $x=2s$, the function $\Gamma(-2s)$ can be rewritten as 
\begin{align}
    \Gamma(-2s)=(-1)^n \frac{\Gamma(n-2s)}{\Gamma(1+2s)}\Gamma(1+2s-n), \hspace{20pt}n\in \mathbb{Z}.
\end{align}
Therefore, for $2s=n+\epsilon$ with $n\in \mathbb{Z}_{\geq 0}$ this function behaves as
\begin{align}
    \Gamma(-2s)\to \frac{1}{\epsilon} \frac{(-1)^{2s}}{\Gamma(2s+1)}\hspace{20pt}
\end{align}
That is, the function $\Gamma(-2s)$ has simple poles when $2s\in \mathbb{Z}_{\geq 0}$, with residue $(-1)^{2s}/\Gamma(2s+1)$. Then, replacing the factor $(-1)^{2s}/\Gamma(2s+1)$ in \eqref{cuasifinal} by $\Gamma(-2s)$ we get a function that has the correct simple poles and residues, and which is valid for any $\alpha$. The final expression for the disk 1-point function (\ref{U}) reads
\begin{equation}
\text{U}_b(\alpha)\,=\, \frac{2}{(2\alpha -Q)}{ \Gamma\Big(2\alpha b-b^2\Big)}{\Gamma\Big(\frac{2\alpha}{b} - \frac{1}{b^2} \Big)}\,\Big( \frac{\pi \Lambda \Gamma(b^2)}{\Gamma(1-b^2)}\Big)^{\frac{Q-2\alpha }{2b }}\cosh\Big(\pi \gamma (2\alpha -Q)\Big)\label{TheU}
\end{equation}
which exactly reproduces the disk 1-point function of spacelike Liouville theory on the disk\footnote{See Eqs. (2.24)-(2.25) in \cite{FZZ}.} \cite{FZZ}. %The change (\ref{charge}) follows from using the property of the $\Gamma$-function in (\ref{properties}) and excluding the pole to extract the residue.
This gives the right pole structure; the 1-point function has poles at
\begin{equation}
\alpha = \frac{b}{2}-\frac{n-1}{2b} \ \ \ \ \ \text{and}\, \ \ \ \ \alpha = -\frac{nb}{2}+\frac{1}{2b} 
\ \ \ \ \ \text{with}\, \ \ n\in\mathbb{Z}_{\geq 0}
\end{equation}
Moreover, (\ref{TheU}) satisfies the reflection relation
\begin{equation}
\text{U}_b(\alpha ) \, =\,  \text{D}(\alpha)\, \text{U}_b(Q-\alpha )\label{reflection}
\end{equation}
where $\text{D}(\alpha)$ is the coefficient that appears in the 2-point function on the sphere and relates the three point functions $C(\alpha_1,\alpha_2,\alpha_3)$ and $C(Q-\alpha_1,\alpha_2,\alpha_3)$; namely,
\begin{equation}
\text{D}(\alpha)\, = \,-\frac{1}{(Q-2\alpha )^2}\Big(\frac{\pi \Lambda \Gamma(b^2)}{\Gamma(1-b^2)}\Big)^{\frac{Q-2\alpha}{b}}\frac{\Gamma(2\alpha b - b^2)\,\,\Gamma(2\alpha /b -1/b^2)}{\Gamma(1-2\alpha b + b^2)\Gamma(1-2\alpha /b +1/b^2)}
\end{equation}
 which reduces to 1 when $\alpha=Q/2$, where $\text{U}_b (\alpha)$ diverges. The reflection property (\ref{reflection}) follows from the fact that the factor $\cosh(\pi \gamma(2\alpha-Q))$  in (\ref{TheU}) is an even function. This simple observation is at the root of the discrepancy\footnote{See the discussions around Eqs. (2.18)-(2.20), and (5.15) in \cite{Anninos}, and  Eqs. (3.29) and (4.2) in \cite{Bautista}.}. between the timelike result proposed in \cite{Anninos} and the one derived in \cite{Bautista}.  

Another interesting quantity is the fixed-length 1-point function, $W_{\ell}(\alpha )$, which is defined from $\text{U}_b(\alpha )$ through the transform
\begin{equation}
    \text{U}_b(\alpha )\, =\, \int _{\mathbb{R}_{\geq 0}}\frac{d \ell }{\ell }\, e^{-\ell \, \Lambda_B } \, W_{\ell }(\alpha )\label{Laplace}
\end{equation}
whose anti-transform reads
\begin{equation}
W_{\ell }(\alpha) \, =\, \frac{\ell }{2\pi i}\, \int_{\epsilon +i\mathbb{R }}d\Lambda_B\, e^{+\ell \, \Lambda_B } \, \text{U}_b(\alpha )\label{antiLaplace}
\end{equation}
with $\epsilon \in \mathbb{R}$ being appropriately defined in terms of the pole structure. Here, $\ell$ represents the length of the perimeter of the disk. Integrating the expression above, we get
\begin{equation}
W_{\ell }(\alpha) \, =\, \frac{2 }{b }\,
\left( \frac{\pi \Lambda \Gamma (b^2)}{\Gamma (1-b^2)}  \right)^{\frac{Q-2\alpha }{2b}}\frac{\Gamma(2\alpha b -b^2)}{\Gamma(1+1/b^2-2\alpha /b)}\, K_{\frac{Q-2\alpha}{b}}\Big(\sqrt{{\Lambda \ell^2 }/{\sin (\pi b^2)}}\Big)
\end{equation}
where $K_{\nu }(x)$ is the modified Bessel function of the second kind.

In the following section, we will perform an analogous calculation for the timelike theory, which, as we will see, presents its peculiarities.

\section{Timelike disk 1-point function}

Let us move to consider the timelike Liouville field theory. In this case, the action takes the form
\begin{equation}
S [\phi ; \Lambda , \Lambda_B]=\frac{1}{4\pi }\int_{\Sigma }d^2x\,\sqrt{g}\,\Big(-g^{\mu\nu}\partial_{\mu}\phi\partial_{\nu}\phi+\hat QR\phi+4\pi\Lambda\, e^{2\beta {\phi }}\Big)+B\label{ActionTimelike}
\end{equation}
with
\begin{equation}
B=\frac{1}{2\pi }\int_{\partial\Sigma}d\text{x}\,\sqrt{h}\,\Big(\hat Q K\phi+2\pi\Lambda_B\,e^{\beta \phi}\Big) \label{ActionTimelike2}
\end{equation}
and the new relation
\begin{equation}
\hat Q=\beta -\frac{1}{\beta }.
\end{equation}
In the expressions above, $R$ is the scalar curvature of the 2-dimensional manifold $\Sigma $ provided with a metric $g$; $h$ is the induced metric in the boundary $\partial \Sigma $, and $K$ is its extrinsic curvature. This defines a non-unitary conformal field theory of central charge
\begin{equation}
c_L=1-6\hat{Q}^2\leq 1
\end{equation}

Now, we compute the 1-point function on the disk in the timelike theory. Following the same steps as in the spacelike theory, we have the resonant correlators
\begin{eqnarray}
\Big\langle \,V_{\alpha}(z, \bar z)\,\Big\rangle\, &=&\, \sum_{m=0}^{\infty } \sum_{l=0}^{\infty } \frac{(-\Lambda )^m}{m!}\frac{(-\Lambda_B )^l}{l!} \, \prod_{i=1}^{m}\int d^2w_i\,
\prod_{k=1}^{l}\int dx_k\,\,\int\mathcal{D}\phi \, \Big( e^{-S[\phi;0,0]}\, e^{2\alpha \phi (z,\bar z )}\,  \nonumber \\ 
&&\ \ \ \  \prod_{i=1}^{m}e^{2\beta \phi(w_i,\bar w_i)}\prod_{k=1}^{l}e^{\beta \phi(x_k)} \, \Big) \, \times \, \delta_{\hat s -m-\frac l2}
\end{eqnarray}
where now
\begin{equation}
\frac{\hat Q-2\alpha}{2\beta }-m-\frac{l}{2}\, =\, 0\ \, , \ \ \ \hat s = \frac 12 -\frac{1}{2\beta ^2}-\frac{\alpha }{\beta }\, . \label{charge2}
\end{equation}
We can choose $z=iy$ with $y\in \mathbb{R}$ and use the timelike free field correlator
\begin{equation}
\langle \phi(z,\bar z ) \phi(w,\bar w ) \rangle \, = \,+\log |z-w||\bar z - w| 
\end{equation}
to obtain the following expression
\begin{eqnarray}
&& \Big\langle V_{\alpha}(z, \bar z)\Big\rangle = \sum_{m=0}^{\infty } \sum_{l=0}^{\infty } \frac{(-1)^{m+l}\Lambda ^m \,\Lambda_B^l}{|2y|^{-2\alpha^2}\, m!\, l!} \prod_{i=1}^{m}\int d^2w_i
\prod_{k=1}^{l}\int dx_k \Big[ \Big( \prod_{k-1}^l (y^2+x_k^2)\prod_{i=1}^{m}|y^2+w_i^2|^2\Big)^{2\alpha \beta }
\nonumber \\ 
&&\ \ \ \ \Big( \prod_{i=1}^{m} \prod_{k=1}^{l} |w_i-x_k|^2 \prod_{i'=1}^m\prod_{i=1}^{i'-1} |w_i-w_{i'}|^2 
\prod_{i'=1}^m\prod_{i=1}^{m} |w_i-\bar{w}_{i'}|
\prod_{k'=1}^l\prod_{k=1}^{k'-1} |x_k-x_{k'}|
\Big)^{2\beta ^2} \Big] \, \times \, \delta_{\hat s -m-\frac l2} \nonumber
\end{eqnarray}
The change in the sign of the exponent of some factors in the integrand changes the pole structure of this integral relative to that of the spacelike case. Still, the integral can be solved by adapting the formulae employed above. The integral is now of the form
\begin{eqnarray}
\hat{\mathcal{J}}_{n,m}(a|y)&=&\frac{1}{m!(n-2m)!}\int \prod_{i}^m d^2w_i\, |y^2+w_i^2|^{-2a}\int \prod_{k=1}^{n-2m}dx_k\, (y^2+x_k^2)^{-a}\, \Big( \prod_{i=1}^m \prod_{k=1}^{n-2m} |w_i-x_k|^2\nonumber \\
&& \ 
\prod_{i=1}^{m}\prod_{i'=1}^{i-1} |w_i-w_{i'}|^2
\prod_{i=1}^{m}\prod_{i'=1}^{m} |w_i-\bar{w}_{i'}|
\prod_{k=1}^{n-2m}\prod_{k'=1}^{k-1} |x_k-x_{k'}|\Big)^{2\beta ^2}
\end{eqnarray}
and can be obtained from its spacelike analog by replacing $b\to i\beta $ and $\alpha \to i \alpha $. That is,
\begin{eqnarray}
&&\hat{\mathcal{J}}_{n,m}(a|y)= \frac{|2y|^{n(1-2a+(n-1)\beta^2)}}{\Gamma(n+1)}\Big(\frac{2\pi }{\Gamma(1+\beta^2)}\Big)^{n}\Big(\frac{1}{4\sin(\pi \beta^2)}\Big)^m\, \times \nonumber\\
&& \ \ \ \ \ \ \ \ \ \ \ \ \ \ \ \ \prod_{i=1}^{n-1}\frac{\Gamma(1+(i+1)\beta^2)\Gamma(2a-1-(n-1+i)\beta^2)}{\Gamma^2(a-i\beta^2)}\, \times  \\
&&\sum_{i=1}^m\frac{(-1)^i \Gamma (n-m-i+1)}{\Gamma (m-i+1)\Gamma(n-2m+1)} \frac{\sin(\pi \beta^2(2i-n-1))}{\sin(\pi \beta^2(i-n-1))}\prod_{r=0}^{i-1}\frac{\sin(\pi \beta^2(r-n))\sin(\pi (a+(r-n)\beta^2))}{\sin(-\pi \beta^2(r+1))\sin(\pi (a-r\beta^2))}\nonumber
\end{eqnarray}
Now, we have $a=-2\alpha \beta $, with $l=n-2m$, $n=2\hat s=l+2m$; in particular, this implies $n(1-2a-(n-1)\beta^2)=2\alpha \hat Q - 4\alpha^2$. Then, repeating the procedure above, we obtain
\begin{equation}
\Big\langle \,V_{\alpha}(iy, -iy)\,\Big\rangle = {|2y|^{2\alpha (\hat Q-\alpha )}}\,2 (-1)^{\hat s}\,\frac{{ \,\Gamma\Big(2\hat{s}\beta ^2+1  \Big)}}{{\Gamma\Big(2\hat{s}+ 1\Big)}}\,\Big(-\pi\Lambda \frac{\Gamma(-\beta ^2)}{\Gamma(1+\beta ^2)}\Big)^{\hat{s}}\,\cosh\Big(2\pi \hat{\gamma} \beta \hat{s}\Big)
\end{equation}
where
\begin{equation}
\cosh^2(\pi \beta  \hat{\gamma} )= -\frac{\Lambda_B^2}{\Lambda }\sin(\pi \beta ^2).\label{hatgamma}
\end{equation}
Using $\hat s = 1/2 -1/(2\beta ^2)-\alpha /\beta $, this yields 
\begin{equation}
\Big\langle \,V_{\alpha}(z, \bar z)\,\Big\rangle = \frac{ 2\, e^{ \frac{-i\pi(\hat Q-2\alpha )}{2\beta}}}{{\big|z-\bar z \big|^{2\alpha (\alpha -\hat Q)}}}\frac{{ \Gamma\Big(\beta ^2-2\alpha \beta  \Big)}}{{\Gamma\Big(2-\frac{2\alpha}{\beta } - \frac{1}{\beta ^2}\Big)}}\Big(  -\frac{\pi \Lambda\Gamma(-\beta ^2)}{\Gamma(1+\beta ^2)}\Big)^{\frac{\hat Q -2\alpha }{2\beta  }}\cosh\Big(\pi \hat{\gamma} (\hat Q - 2\alpha )\Big)\label{resulttimelike}
\end{equation}
where we have rewritten the factor $(-1)^{\hat s}$ by $e^{-i\pi(\hat Q-2\alpha )/(2\beta)}$ in anticipation to the discussion in section \ref{section reflection}.

As we see, the solution (\ref{resulttimelike}) for the disk 1-point function for the timelike theory exhibits the factor $\text{cosh}(\pi \hat{\gamma} (\hat Q - 2\alpha ))$, which is an even function of the variable defined in (\ref{hatgamma}) in terms of the ratio of the cosmological constants $\Lambda^2_B/\Lambda$.

Before continuing, let us make some remarks on the relation between the spacelike and the timelike 1-point function and the integration over the zero-mode. In \cite{Giribet}, a derivation of the 3-point function of the timelike theory on the sphere using the Coulomb gas formalism was presented. The result obtained there was consistent with the timelike version of the Dorn-Otto-Zamolodchikov-Zamolodchikov (DOZZ) formula\cite{Zamolodchikov:2005fy}; that is, an expression that, up to a normalization of the vertices, can be considered as the {\it inverse} of the naive analytic continuation one would propose for such an observable from its spacelike counterpart, cf. \cite{ZZ}. An interesting observation arising from the derivation presented in \cite{Giribet} is that, in the case of the timelike theory, an additional divergent factor appears; the latter is expressed by a factor of $\Gamma(0)$. After removing this factor, the result matches exactly that proposed in the literature \cite{HMW}. However, the explanation for the appearance of such a factor in the timelike case remained in \cite{Giribet} an intriguing open question and marks an important distinction with respect to the spacelike case --where such extra divergence does not appear. The difference between both theories is presumably due to the way in which the integration of the zero-mode must be treated in the timelike theory \cite{Santachiara}. In the timelike theory, integrating the zero-mode {\it ab initio} does not seem to lead to the correct result. In other words, the calculation in the timelike theory requires one to extract the extra divergence appearing in the naive integration of the zero-mode. Keeping this in mind, we propose that the timelike 1-point function is simply given by $\langle V_{\alpha}(z, \bar z)\rangle = |z-\bar z |^{2\alpha (\hat Q-\alpha )}\hat{\text{U}}_{\beta}(\alpha)$ with
%\begin{equation}
%\hat{\text{U}}_{\beta}(\alpha )\, =\, \frac{4\pi }{\beta} \,e^{- \frac{i\pi(\hat Q-2\alpha )}{2\beta}}\,
%\frac{{ \Gamma\Big(\beta ^2-2\alpha \beta  \Big)}}{{\Gamma\Big(2-\frac{2\alpha}{\beta } - \frac{1}{\beta ^2}\Big)}}\left(  -\frac{\pi \Lambda\Gamma(-\beta ^2)}{\Gamma(1+\beta ^2)}\right)^{\frac{\hat Q -2\alpha }{2\beta  }}\cosh\Big(\pi \hat{\gamma} (\hat Q - 2\alpha )\Big)\label{resulttimelikes}
%\end{equation}
\begin{equation}
\hat{\text{U}}_{\beta}(\alpha )\, =\, \frac{4\pi }{\beta} e^{-i\pi\frac{\hat{Q}-2\alpha}{2\beta}}\,
\frac{{ \Gamma\Big(\beta ^2-2\alpha \beta  \Big)}}{{\Gamma\Big(2-\frac{2\alpha}{\beta } - \frac{1}{\beta ^2}\Big)}}\left(-  \frac{\pi \Lambda\Gamma(-\beta ^2)}{\Gamma(1+\beta ^2)}\right)^{\frac{\hat Q -2\alpha }{2\beta  }}\cosh\Big(\pi \hat{\gamma} (\hat Q - 2\alpha )\Big)\label{resulttimelikes}
\end{equation}
with $\hat{Q}=\beta-1/\beta $ and $\hat \gamma $ being defined as $\cosh^2(\pi \beta  \hat{\gamma} )= -({\Lambda_B^2}/{\Lambda })\sin(\pi \beta ^2)$. 
The normalization ${2\pi }/{\beta}$, which comes from the integration over the zero mode --see the length of $\mathcal{C}_R^{(2)}$ in section \ref{section41}-- has the right properties as will be seen in the following. Formula (\ref{resulttimelikes}) is the main result of this paper. Notice that this differs from the result in \cite{Bautista}, especially in the dependence on the parameter\footnote{See Eq. (3.29) in \cite{Bautista}, where $s$ there is $\hat\gamma $ here, $\mu $ there is $\Lambda $ here. Cf. Eq. (\ref{resulttimelikesss}) below.} $\hat \gamma$.

As a first observation, it is worth noting the curious relation this result has with the analytical extension of the spacelike case (\ref{TheU}). For $\hat s \notin \mathbb{Z}$, we find the following relation
\begin{equation}
 \,\Big\langle V_{\alpha}(z,\bar z) \Big\rangle_{\substack{\text{\,Timelike}\\\text{LFT $\beta$}}}\, =-2i\,{e^{-2i\pi\hat s}\sin (2\pi \hat s)} \,\times \,\Big\langle V_{i\alpha}(z,\bar z) \Big\rangle_{\substack{\text{\,Spacelike}\\\text{\,  LFT $i\beta$}}}  \label{casiresemblance}
\end{equation}
which, denoting $s=2\hat{s}=(\hat{Q}-2\alpha) /\beta $, can be written as
\begin{equation}
\Big\langle V_{i\alpha}(z,\bar z) \Big\rangle_{\substack{\text{\,Spacelike}\\\text{\,  LFT $i\beta$}}} \, =\,\, \frac{i\,e^{i\pi s}}{2\sin (\pi s)}\,\, \, \,\Big\langle V_{\alpha}(z,\bar z) \Big\rangle_{\substack{\text{\,Timelike}\\\text{LFT $\beta$}}}\,
\label{resemblance}
\end{equation}
This is exactly the same relation between the timelike and the extended spacelike 1-point function found for the case $\Lambda=0$ in \cite{Santachiara}. We will make the relation to the derivation in \cite{Santachiara} explicit in the next sections: we will see that our result for $\hat{\text{U}}_{\beta}(\alpha )$ above reduces to that of \cite{Santachiara} in the limit $\Lambda \to 0$. Moreover, in the next section we will also perform several consistency checks of our formula (\ref{resulttimelikes}), such as showing that it has the right reflection symmetry and satisfies the bootstrap shift-equations.

\section{Consistency checks}
\subsection{Timelike disk 1-point function at $\Lambda =0$}\label{section41}

We will first review the calculation in \cite{Santachiara} and, after that, show that the $\Lambda \to 0$ limit of (\ref{resulttimelikes}) agrees with it. In order to facilitate the comparison, we will parallel the approach in \cite{Santachiara}, which in particular means keeping the original spacelike field $\phi $ but taking the imaginary values for $b\to i\beta $ and $\alpha \to i\alpha$; also, as in \cite{Santachiara}, we set $\Lambda =0$. This yields the action\footnote{More precisely, the notation in \cite{Santachiara} relates to ours as follows: $\beta $ there is $2\beta $ here, and $\mu_B$ there is $2\pi \Lambda_B$ here.}
\begin{equation}
S [\phi ; 0 , \Lambda_B]=\frac{1}{4\pi }\int_{\Sigma }d^2x\,\sqrt{g}\,\Big(g^{\mu\nu}\partial_{\mu}\phi\partial_{\nu}\phi+i\hat QR\phi\Big)+B
\end{equation}
with the boundary term
\begin{equation}
B=\frac{1}{2\pi }\int_{\partial\Sigma}d\text{x}\,\sqrt{h}\,\Big(i\hat QK\phi+2\pi\Lambda_B\,e^{i\beta \phi}\Big)
\end{equation}
and the relation $\hat Q=\beta-1/\beta$. We also have $\Delta_{\alpha} = \alpha (\alpha -\hat Q)$. 

The 1-point function is defined as
\begin{equation}
\Big\langle \,e^{2i\alpha \phi(z,\bar z)}\,\Big\rangle\, =\, \int\mathcal{D}\phi \, e^{-S[\phi ;0, \Lambda_B]}\, e^{2i\alpha \phi (z,\bar z )}\, =\, 
\int_{\mathcal{C}}dc\,e^{ic(2\alpha-\hat Q)}\int\mathcal{D}\varphi \, e^{-S[\varphi ;0, \Lambda_B e^{i\beta c}]}e^{2i\alpha \varphi (z,\bar z )}
\end{equation}
where in the second inequality we have separated the integral over the zero-mode $c$ and the fluctuations, according to the following definition: $\phi (z,\bar z)=c+\varphi (z,\bar z)$. We can redefine the boundary cosmological constant as follows $\tilde \Lambda_B = \Lambda_Be^{i\beta c}$. The integration contour $\mathcal{C}$ will be specified later. For the moment, let us consider the part of the integral that involves the fluctuations $\varphi(z,\bar z)$. Consider the expansion
\begin{equation}
\int\mathcal{D}\varphi \, e^{-S[\varphi ;0, \tilde \Lambda_B ]}e^{2i\alpha \varphi (z,\bar z )}=\sum_{l=0}^{\infty}\frac{(-\tilde\Lambda_B)^l}{l!}\int \prod_{k=1}^{l}dx_k\,\Big( \int\mathcal{D}\varphi \, 
e^{-S[\varphi ;0, 0]}e^{2i\alpha \varphi (z,\bar z )}\prod_{k=1}^{l}e^{i\beta \varphi(x_k)}\Big)
\end{equation}
where now the correlator on the right hand side is defined in the free theory with a background charge. Setting $z=0$ and using the spacelike free field propagator (recall that, following \cite{Santachiara}, we did not Wick-rotate the field $\phi$), we obtain
\begin{equation}
\int\mathcal{D}\varphi \, e^{-S[\varphi ;0, \tilde \Lambda_B ]}e^{2i\alpha \varphi (0)}=\sum_{l=0}^{\infty}\frac{(\tilde\Lambda_B)^l}{l!}\int \prod_{k=1}^{l}dx_k\,\prod_{k=1}^{l}\prod_{k'=1}^{k-1}|x_k-x_{k'}|^{2\beta^2}
\end{equation}
Now, we can closely follow the analysis in \cite{Santachiara}. Using formula (36) therein, we have
\begin{equation}
\int \prod_{i=1}^{n}d\theta_i \prod_{k=1}^{n}\prod_{j=1}^{k-1} \large|e^{i\theta_j}-e^{i\theta_k}\large|^{2\beta^2} = {(2\pi )^n}\, \frac{\Gamma (1+n\beta^2)}{(\Gamma (1+\beta^2))^n}  
\end{equation}
which, using $\Gamma(1+l\beta^2)=\int_{\mathbb{R}>0}t^{l\beta^2}e^{-t}dt$, leads to 
\begin{equation}
\sum_{l=0}^{\infty } \Big(\frac{-2\pi \tilde \Lambda_B}{\Gamma(1+\beta^2)}\,\Big)^l\,\,\frac{\Gamma(1+l\beta^2)}{{\Gamma(l+1)}}\, =\, \int_{\mathbb{R}_{> 0}}dt\, e^{-\frac{2\pi \tilde\Lambda_B t^{\beta^2}}{\Gamma(1+\beta^2)}-t}
\end{equation}
Then, we have
\begin{equation}
\Big\langle \,e^{2i\alpha \phi(0)}\,\Big\rangle\, =\,  
\int_{\mathcal{C}}dc\,e^{ic(2\alpha-\beta+\frac{1}{\beta})}\, \int_{\mathbb{R}_{> 0}}dt\, e^{-\frac{2\pi \Lambda_B e^{i\beta c } t^{\beta^2}}{\Gamma(1+\beta^2)}-t}
\end{equation}
The contour $\mathcal{C}$ in the complex $c\in \mathbb{C}$ has been prescribed in \cite{Santachiara}. It is defined as the union of two semi-infinite lines and a segment. More precisely, we have $\mathcal{C}=\mathcal{C}^{(1)}\cup \mathcal{C}^{(2)}\cup \mathcal{C}^{(3)}$, where $\mathcal{C}^{(1)}=\{c\in \mathbb{C} \,|\,c=i\tau \text{ with }\tau \in (-\infty , 0)\}$, $\mathcal{C}^{(2)}=\{c\in \mathbb{C} \,|\,c=\tau \text{ with }\tau \in (0 , 2\pi)\}$, and $\mathcal{C}^{(3)}=\{c\in \mathbb{C} \,|\,c=2\pi -i\tau \text{ with }\tau \in (0 , \infty)\}$. This is depicted in Figure 2. 
\begin{figure}[h]
    \begin{center}
        \includegraphics[width=14.0
        cm]{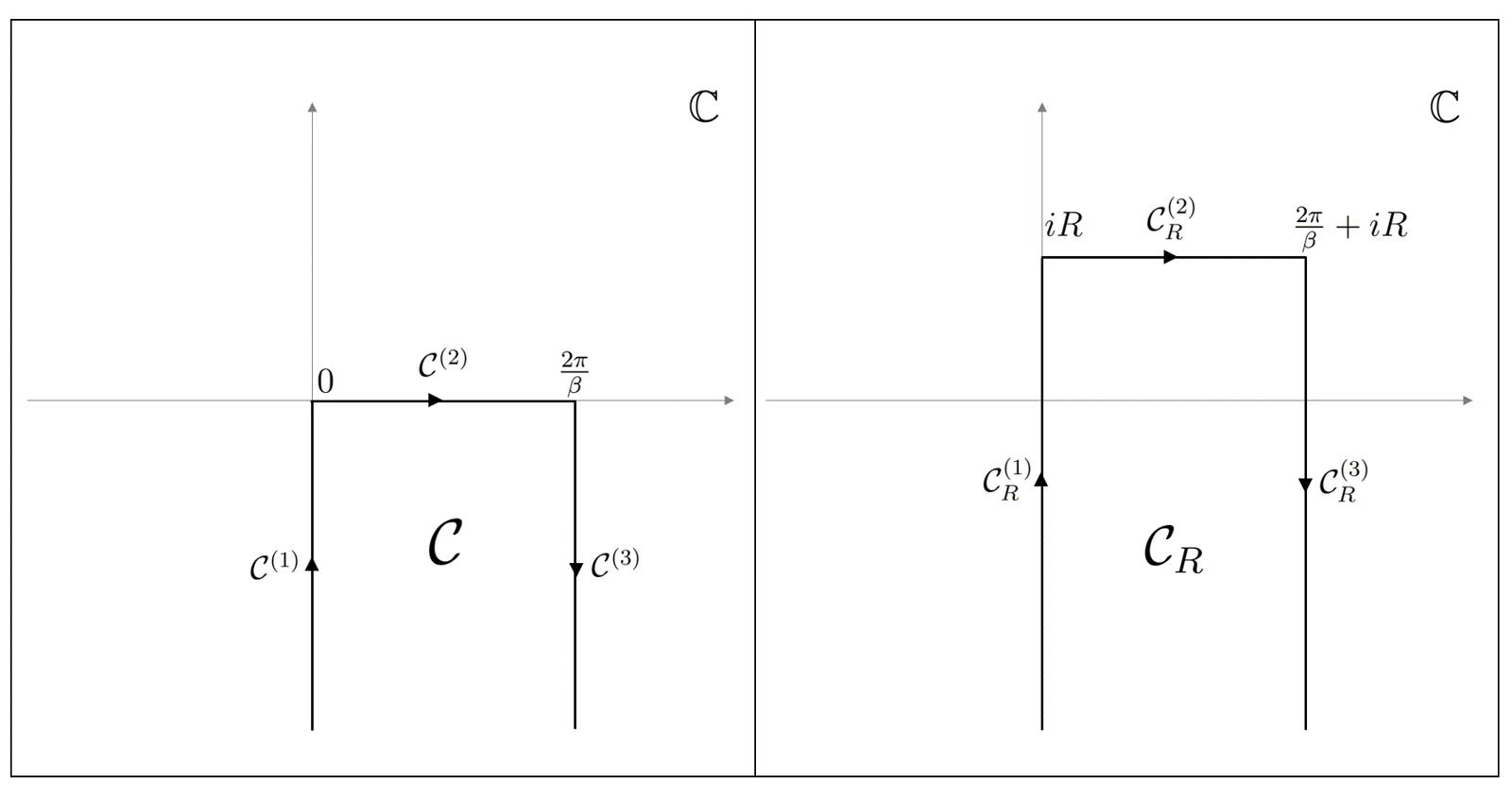}
        \caption{Integration contours $\mathcal{C}$ and $\mathcal{C}_R$ in the $c\in \mathbb{C}$ plane.}
        \label{Figure2}
    \end{center}
\end{figure}
The calculation can be performed by deforming the contour as follows $\mathcal{C}\to \mathcal{C}_R$, with $\mathcal{C}_R=\mathcal{C}_R^{(1)}\cup \mathcal{C}_R^{(2)}\cup \mathcal{C}_R^{(3)}$, where $\mathcal{C}_R^{(1)}=\{c\in \mathbb{C} \,|\,c=i\tau \text{ with }\tau \in (-\infty , R)\}$, $\mathcal{C}_R^{(2)}=\{c\in \mathbb{C} \,|\,c=\tau +iR \text{ with }\tau \in (0 , 2\pi)\}$, and $\mathcal{C}_R^{(3)}=\{c\in \mathbb{C} \,|\,c=2\pi -i\tau \text{ with }\tau \in (-R , \infty)\}$, with $R\in \mathbb{R}$, and then taking $R\to \infty$. By doing so, one finds the following
\begin{equation}
\Big\langle \,e^{2i\alpha \phi(0)}\,\Big\rangle\, =\, i\,\Big(1-e^{-2\pi is}\Big)\, \int_{\mathbb{R}_{> 0}}dt\, e^{-t}\, \int_{\mathbb{R}}dx\, e^{x(\alpha -\beta+\frac{1}{\beta})}\,e^{-\frac{2\pi \Lambda_Be^{\beta x}t^{\beta^2}}{\Gamma(1+\beta^2)}}
\end{equation}
with $s=1-1/\beta^{2}-2\alpha /\beta$. Changing variables $y:=e^{\beta x}$, integrating in $y$ and using the integral representation of the $\Gamma$-function, $\int_{\mathbb{R}_{>0}}dt\,t^{x-1}e^{-t}=\Gamma(x) $, we finally obtain
\begin{equation}
\Big\langle \,e^{2i\alpha \phi(0)}\,\Big\rangle\, =\, \frac{2\pi }{\beta}\,e^{-i\pi s}\,\Big(\frac{2\pi \Lambda_B}{\Gamma(1+\beta^2)}\Big)^s\,\frac{\Gamma(1+s\beta^2)}{\Gamma(1+s)}\label{finalisima}
\end{equation}
where the properties of the $\Gamma$-function written above, like $\Gamma(-s)\Gamma(1+s)=-\pi/\sin(\pi s)$, were of use. Expression (\ref{finalisima}) is exactly the expression obtained in \cite{Santachiara}; see Eqs. (8)-(11) therein. This shows the correct way of treating the zero-mode in the timelike theory.

\subsection{Comparison with our result}

Now, let us compare this result with the $\Lambda \to 0$ limit of our formula (\ref{resulttimelikes}). This limit is subtle, as $\Lambda $ enters in the definition of $\hat \gamma $ in (\ref{hatgamma}). If one performs the limit carefully, in which one observes that $e^{2\pi \beta \hat \gamma}\simeq -(\Lambda_B^2/\Lambda )\sin(\pi \beta )$ when $\Lambda/\Lambda_B^2\to 0$, and assumes the timelike version of the Seiberg bound $\hat Q>2\alpha $, one obtains that (\ref{resulttimelike}) actually yields
\begin{equation}
\lim_{\Lambda \to 0}\hat{\text{U}}_{\beta}(\alpha) \, = \frac{2\pi }{\beta}\,e^{- i\pi s}\,\left(\frac{2\pi \Lambda_B}{\Gamma(1+\beta^2)}\right)^{s}\,\frac{\Gamma(1+s\beta^2)}{\Gamma(1+s)}\, 
\end{equation}
with $s=(\hat{Q}-2\alpha)/\beta$,
which exactly matches (\ref{finalisima}). This match is essential to set the prefactor $2\pi /\beta $ in the definition of $\hat{\text{U}}_{\beta}(\alpha )$ above. It is remarkable that the non-trivial part of this formula, i.e., the $s$-dependent $\Gamma$-functions, matches exactly. In the next subsection we will see that the Coulomb gas realization which assumes $\Lambda =0$ from the beginning reproduces exactly the same result. 

Interestingly, when the Seiberg bound is not obeyed, i.e., when $\hat Q<2\alpha $, taking the limit $\Lambda\to 0$ of (\ref{resulttimelike}) by keeping $\Lambda_B$ fixed leads to a divergent result. However, there is a way to take that limit and reproduce (\ref{finalisima}) even for $\hat{Q}<2\alpha $. To achieve that, one has to take the double scaling limit $\Lambda\to 0$, $\Lambda_B \to 0$ keeping the ratio 
\begin{align}\label{alternative limit}
    \frac{\Lambda}{\Lambda_B}\,:=\,-4\Lambda_B' \sin(\pi \beta^2)
\end{align}
fixed, with the new boundary cosmological constant $\Lambda_B'$ being defined by this relation. In this limit, (\ref{resulttimelike}) reduces to (\ref{finalisima}) with the boundary constant $\Lambda_B \to \Lambda'_B$ for $\hat{Q}<2\alpha $.

In order to understand the latter limit, it is instructive to compare it with the spacelike case: {In fact, the necessity of taking the limit carefully when $2\alpha >\hat{Q}$ also appears in that case. Using the Coulomb gas formalism to compute the spacelike 1-point function with $\Lambda =0$, integrating the zero-mode over the real line at the beginning of the computation, then solving the Selberg type integrals, and finally extending the result for $s=(Q-2\alpha)/b\in \mathbb{C}$, yields
\begin{align}
    \text{U}_{b}(\alpha)_{\Lambda=0}\, =\, \frac{1}{b}\left(\frac{2\pi \Lambda_B}{\Gamma(1-b^2)}\right)^{\frac{Q-2\alpha}{b}}\Gamma(2\alpha b-b^2)\Gamma\left(\frac{2\alpha}{b}-\frac{1}{b^2}-1\right).
\end{align}
The result for the 1-point function with $\Lambda \neq 0$ (\ref{TheU}) tends to this expression in the limit $\Lambda \to 0$ if $2\alpha<Q$. In contrast, if $2\alpha>Q$, the limit $\Lambda\to 0$ while keeping $\Lambda_B$ fixed diverges, and to match the $\Lambda=0$ result, one has to take the double scaling limit $\Lambda\to0$, $\Lambda_B\to 0$ while keeping fixed ${\Lambda}/{\Lambda_B}=4\Lambda_B'\sin(\pi b^2)$.

\subsection{Resonant correlators at $\Lambda =0$}

Now, in order to fully confirm that the Coulomb gas computation yields the correct result, let us compute the resonant correlator by setting $\Lambda =0$ from the beginning --instead of taking the $\Lambda \to 0 $ in (\ref{resulttimelikes})-- and then compare with (\ref{finalisima}). This amounts to inserting boundary screening operators to saturate the charge condition $2\alpha +s\beta =\hat Q$, namely,
\begin{eqnarray}
\Big\langle \,e^{2\alpha \phi (0)}\,\Big\rangle\, =\,  \frac{(-\Lambda_B)^s}{s! }\int \prod_{k=1}^{s} dx_k\, \int \mathcal{D}\varphi \, e^{-S[\varphi ; 0, 0]}\, e^{2\alpha \varphi (z,\bar z )}\, \prod_{k=1}^{s}e^{\beta \varphi (x_k)}
\end{eqnarray}
This is similar to the computation performed in previous sections, but now considering $\Lambda =0$, so it brings us back to an expression like 
\begin{eqnarray}
\Big\langle \,e^{2\alpha \phi (0)}\,\Big\rangle\, =\,  \frac{(-\Lambda_B)^s}{s!}\int \prod_{k=1}^{s} dx_k\, \prod_{k=1}^{s}|x_k|^{4\alpha \beta }\prod_{k=1}^{s}\prod_{k'=1}^{k-1}|x_k-x_{k'}|^{2\beta^2}\,
\end{eqnarray}
which yields
\begin{eqnarray}
\Big\langle \,e^{2\alpha \phi (0)}\,\Big\rangle\, =\,   e^{-i\pi s}\, \Big(\frac{2\pi \Lambda_B}{\Gamma(1+\beta^2)}\Big)^s\,\, \frac{\Gamma(1+s\beta^2)}{\Gamma(1+s)}\nonumber
\end{eqnarray}
This times the factor $2\pi /\beta$ exactly matches (\ref{finalisima}) . The latter factor appears in the integration over the zero-mode following \cite{Santachiara} and coincides with the length of the segment $\mathcal{C}_{R}^{(2)}$. {This is the same factor that we add to the computation of the $\Lambda\neq0$ case in section 3 to get to the final result (\ref{resulttimelikes})}. In summary, the Coulomb gas computation of the resonant correlators suffices to reproduce the correct result for the timelike 1-point function. 

\subsection{The zero-mode: a cautionary note}

In this subsection, we will discuss the integration over the zero-mode in the timelike theory, which, as observed in \cite{Giribet}, exhibits a difference with respect to the spacelike theory. What we are going to show is that, as happens in the case of the DOZZ formula, if one first integrates the timelike zero-mode instead of proceeding as we did above, then the result receives an extra divergent factor $\sim \Gamma(0)$ that has to be extracted. This is a feature of the timelike theory that does not occur in the spacelike theory.

First, let us go back to our notation in (\ref{ActionTimelike})-(\ref{ActionTimelike2}), where $\phi $ is a timelike field and the action is real. Then, the 1-point function reads
\begin{eqnarray}
\Big\langle \,e^{2\alpha \phi (z,\bar z)}\,\Big\rangle\, =\, \int\mathcal{D}\phi \, e^{-S[\phi ;0 , \Lambda_B]}\, e^{2\alpha \phi (z,\bar z )}\, &=& \, \int _{\mathbb{R}_{>0}}dA\, e^{-\Lambda_BA}\, \int \mathcal{D}\varphi\, e^{-S[\varphi ; 0,0]}\, e^{2\alpha \varphi(z,\bar z )} \nonumber \\ 
&&  \times \, \int_{\mathbb{R}}dc\, e^{c( 2\alpha - \hat Q)}\,
\delta\Big(A-e^{c\beta}\int dx\, e^{\beta\varphi(x)}\Big)\label{f}
\end{eqnarray}
where, as before, $\phi (z,\bar z)=c+\varphi (z,\bar z)$. Using properties of the $\delta$-function, we get
\begin{eqnarray}
\Big\langle \,e^{2\alpha \phi (z,\bar z)}\,\Big\rangle\, &=& \, \int _{\mathbb{R}_{>0}}dA\, e^{-\Lambda_BA}\, \int \mathcal{D}\varphi\, e^{-S[\varphi ; 0,0]}\, e^{2\alpha \varphi(z,\bar z )} \nonumber \\ 
&&  \times \, \int_{\mathbb{R}}\frac{dc}{\beta A}\, e^{c( 2\alpha - \beta +\frac{1}{\beta })}\,
\delta\Big(c+\frac{1}{\beta }\log \Big( \frac 1A \int dx\,e^{\beta \varphi (x)}\Big)\Big)\, 
\end{eqnarray}
Now, we will first integrate the zero-mode to see what happens: by integrating $c$, we obtain \cite{GL}
\begin{eqnarray}
\Big\langle \,e^{2\alpha \phi (z,\bar z)}\,\Big\rangle\, &=&\,   \frac{1}{\beta}\int_{\mathbb{R}_{>0}}\frac{dA}{A}\, e^{-\Lambda_B A}\, \int\mathcal{D}\varphi \, e^{-S[\varphi ; 0, 0 ]}\, e^{2\alpha\varphi(z,\bar z)} \, \Big(\frac 1A\, \int dx\, e^{\beta\phi(x)}\Big)^{1-\frac{1}{\beta^2}-\frac{2\alpha}{\beta}}
\end{eqnarray}
Then, using $\int_{\mathbb{R}_{>0}}dA\, A^{-s-1} e^{-\Lambda_BA}=\Lambda_B^{s}\Gamma(-s)$ and $s=1-1/\beta^2-2\alpha /\beta$, we obtain the following
\begin{eqnarray}
\Big\langle \,e^{2\alpha \phi (z,\bar z)}\,\Big\rangle\, =\,  \frac{\Lambda^s\Gamma(-s)}{\beta }\int \mathcal{D}\varphi \, e^{-S[\varphi ; 0, 0]}\, e^{2\alpha \varphi (z,\bar z )}\, \Big(\int dx\, e^{\beta \varphi (x)}\Big)^s
\end{eqnarray}
which is nothing but the formal expression for the multiple correlator
\begin{eqnarray}
\Big\langle \,e^{2\alpha \phi (z,\bar z)}\,\Big\rangle\, =\,  \frac{\Lambda_B^s\Gamma(-s)}{\beta }\int \prod_{k=1}^{s} dx_k\, \int \mathcal{D}\varphi \, e^{-S[\varphi ; 0, 0]}\, e^{2\alpha \varphi (z,\bar z )}\, \prod_{k=1}^{s}e^{\beta \varphi (x_k)}
\end{eqnarray}
provide we assume $s\in\mathbb{Z}_{\geq 0}$. Using the timelike free propagator and setting $z=0$, we can write this as follows
\begin{eqnarray}
\Big\langle \,e^{2\alpha \phi (0)}\,\Big\rangle\, =\,  \frac{\Lambda_B^s\Gamma(-s)}{\beta }\int \prod_{k=1}^{s} dx_k\, \prod_{k=1}^{s}|x_k|^{4\alpha \beta }\prod_{k=1}^{s}\prod_{k'=1}^{k-1}|x_k-x_{k'}|^{2\beta^2}\, .
\end{eqnarray}
This integral is a particular case of the one solved in the previous sections. Using the same formulae and properties of the $\Gamma$-functions (\ref{properties}), we obtain
\begin{eqnarray}
\Big\langle \,e^{2\alpha \phi (0)}\,\Big\rangle\, =\,  \frac{\Lambda_B^s\Gamma(-s)}{\beta }\, \Gamma(1+ s \beta^2)\, \Big(\frac{2\pi \Lambda_B}{\Gamma(1+\beta^2)}\Big)^{s}
\end{eqnarray}
Now, we can multiply and divide by $\Gamma(1+s)$ and recall that we have assumed $s\in \mathbb{Z}_{\geq 0}$, so that $\Gamma(-s)$ yields a divergence. Rewriting this divergent factor using $\Gamma(-s)\Gamma(1+s)=e^{-i\pi s}\Gamma(0)$, we get
\begin{eqnarray}
\Big\langle \,e^{2\alpha \phi (0)}\,\Big\rangle\, =\, \frac{\Gamma(0)}{2\pi }\, \times\, \frac{2\pi \, e^{-i\pi s}}{\beta} \,\Big(\frac{2\pi \Lambda_B}{\Gamma(1+\beta^2)}\Big)^{s}\, \frac{\Gamma (1+s\beta^2)}{\Gamma(1+s)}\label{lafinalisima}
\end{eqnarray}
which, up to a divergent factor $\sim \Gamma (0) $ coincides with (\ref{finalisima}). This is exactly the same phenomenon observed in \cite{Giribet} for the case of the sphere 3-point function. Hence, it allows us to identify where the divergent factor emerges: as suspected, it appears throughout the integration of the zero-mode. Such a divergence does not appear if the integration over the zero-mode is performed following the prescription \cite{Santachiara}. In other words, the prescription for the integration contours proposed in \cite{Santachiara} is equivalent to removing the divergent factor $\Gamma(0)$; i.e., the method in \cite{Santachiara} is the correct way to solve the problem observed in \cite{Giribet}. This exercise has been very instructive; it is what led us to propose our result (\ref{resulttimelikes}).

\subsection{Reflection symmetry}\label{section reflection}

In this subsection, as another cross check of our result, we will prove that formula (\ref{resulttimelikes}) obeys the correct reflection properties. In order to prove that, let us recall the timelike DOZZ formula \cite{HMW}, namely
\begin{equation}
\big\langle\, V_{\alpha_1}(z_1, \bar z_1)\,V_{\alpha_2}(z_2, \bar z_2)\,V_{\alpha_3}(z_3, \bar z_3)\,\big\rangle \, = \, \hat{C}(\alpha_1, \alpha_2, \alpha_3)\, \prod_{i<j}^3 |z_i-z_j|^{2\Delta-4(\Delta_{\alpha_i}+\Delta_{\alpha_j} )}
\end{equation}
with $\Delta:=\Delta_{\alpha_1}+\Delta_{\alpha_2}+\Delta_{\alpha_3}$ and
\begin{equation}
\hat{C}(\alpha_1 , \alpha_2 , \alpha_3 )\, = \, \frac{2\pi}{\beta}e^{-\frac{i\pi (\hat Q -\alpha)}{\beta} }\, \Big(-\pi \Lambda \frac{\Gamma(-\beta^2)}{\Gamma(1+\beta^2)}\beta^{2+2\beta^2}\Big)^{\frac{\hat Q-2\alpha}{\beta}}\, \frac{\Upsilon _{\beta} (\hat Q+\beta-\alpha)}{\Upsilon _{\beta} (\beta )}\, \prod_{i=1}^{2}\frac{\Upsilon _{\beta} (2\alpha_i -\alpha +\beta )}{\Upsilon _{\beta} (\beta -2\alpha _i)}\label{DOZZt}
\end{equation}
with $\alpha :=\alpha_1+\alpha_2+\alpha_3$. See \cite{Zamolodchikov:2005fy} for the definition of the $\Upsilon_{\beta}$-function. This function, discussed in the context of the timelike theory, can also be found in \cite{HMW} and \cite{Giribet}. 

From (\ref{DOZZt}) it is possible to obtain the timelike reflection coefficient $\hat{\text{D}}(\alpha )$, namely
\begin{equation}
\hat{\text{D}}(\alpha _1) \, =\, \frac{\hat{C}(\alpha_1 , \alpha_2 , \alpha_3 )}{\hat{C}(\hat Q -\alpha_1 , \alpha_2 , \alpha_3 )} 
\end{equation}
which yields
\begin{equation}
\hat{\text{D}}(\alpha ) \, = \, -\frac{e^{-\frac{i\pi (\hat Q-2 \alpha)}{\beta}}}{(\hat Q-2\alpha)^2}\, \Big(-\frac{\pi \Lambda \Gamma(-\beta^2)}{\Gamma(1+\beta^2)}\Big)^{\frac{\hat Q-2\alpha}{\beta }} \frac{\Gamma(\beta^2-2\beta \alpha)\Gamma(2\alpha/\beta + 1/\beta^2)}{\Gamma(1-\beta^2+2\beta \alpha )\Gamma(1-2\alpha /\beta - 1/\beta^2)}\, .
\end{equation}
This is the timelike reflection coefficient. The next step is to prove that our result for $\hat{\text{U}}_{\beta}(\alpha )$ is consistent with it. In fact, one can easily show that this is actually the case: using (\ref{resulttimelike}) or (\ref{resulttimelikes}), one finds
\begin{equation}
{\hat{\text{U}}_{\beta}(\alpha)} \, = \, \hat{\text{D}}(\alpha )\, {\hat{\text{U}}_{\beta}(\hat Q -\alpha )}\, .  
\end{equation}
This property has been one of the ingredients in the recent discussion in the quantum cosmology applications; see \cite{Anninos, Bautista}. 

\subsection{Fixed-length 1-point function}

Now, we can proceed to compute the fixed-length 1-point function $   \hat{W}_{\ell }(\alpha )$ in a similar way to the spacelike case. This amounts to computing the inverse-transform (\ref{antiLaplace}) for $\hat{\text{U}}_{\beta}(\alpha)$. The simplest case to start with is $\Lambda =0 $, for which the result of the 1-point function has been obtained in \cite{Santachiara} and has been shown to coincide with the limit $\Lambda \to 0$ of ours. By performing the integral (\ref{antiLaplace}) with (\ref{resulttimelikes}), which amounts to adequately dealing with the integration contour, one finally obtains the following result
\begin{equation}
    \hat{W}_{\ell }(\alpha )\Big|_{\Lambda=0}\, =\, -\frac{2}{\beta }\, e^{-\frac{i\pi(\hat Q -2\alpha)}{\beta }}\, \sin\big(\pi(\hat Q-2\alpha)/\beta \big)\, \Gamma(\beta^2-2\alpha \beta ) \, 
    \left(\frac{2\pi }{\Gamma(1+\beta^2)\ell}\right)^{\frac{\hat Q-2\alpha }{\beta}}
    \label{bdy_length_lambda=0}
\end{equation}
Notice that for $\hat{Q}>2\alpha $ this expression diverges in the limit $\ell \to 0$, while for $\hat{Q}<2\alpha $ this tends to zero in the limit $\ell \to 0$. This parallels the expectation\footnote{See the discussions around Eqs. (2.19), (5.12), (5.27) and (6.2) in \cite{Anninos}} of \cite{Anninos} for $\Lambda >0$.

{The calculation in the case $\Lambda > 0$ is more cumbersome but can still be performed analytically. Deforming the contour as in Figure 3 we arrive at\footnote{This expression is valid for $\hat Q-2\alpha  $ generic enough. Values of $\alpha $ such that $\nu =(\hat Q-2\alpha )/\beta \in \mathbb{Z}$ correspond to confluent points of the Bessel equation, where $J_{-\nu}(x)=(-1)^{\nu }J_{\nu }(x)$. Notice also the presence of the prefactor $\Gamma^{-1}(\nu )$, which vanishes for $\nu\in \mathbb{Z}_{\leq 0}$.}
\begin{align}
    \hat{W}_{\alpha}(\ell)=\frac{2\pi}{\beta} &\frac{\Gamma(\beta^2-2\alpha\beta)}{\Gamma\left(\frac{\hat{Q}-{2\alpha}}{\beta }\right)}\Bigg(-\frac{\pi \Lambda\Gamma(-\beta ^2)}{\Gamma(1+\beta ^2)}\Bigg)^{\frac{\hat Q -2\alpha }{2\beta  }}\nonumber\\ &\times\left[J_{\frac{\hat{Q}-2\alpha}{\beta}}\left(\sqrt{\frac{\Lambda \ell^2}{\sin(\pi \beta^2)}}\right)-e^{-i\pi \frac{\hat{Q}-2\alpha}{\beta}} J_{-\frac{\hat{Q}-2\alpha}{\beta}}\left(\sqrt{\frac{\Lambda \ell^2}{\sin(\pi \beta^2)}}\right)\right].
    \label{bdy length timelike}
\end{align}
Where the $J_{\nu}(x)$ are the Bessel functions of the first kind.} 

{It is worth noting that the result (\ref{bdy length timelike}) does not have the form proposed in \cite{Anninos}, where the authors argued that the fixed length 1-point function should involve a single Bessel function. In \cite{Bautista}, the timelike 1-point function was obtained by analytically continuing the bootstrap equations from the spacelike calculation. This yielded a linear combination of Bessel functions whose relative coefficient remained to be fixed by symmetry arguments. The authors of \cite{Anninos} found the result of \cite{Bautista} at odds with their expectation from the cosmology applications viewpoint. In contrast to the method employed in \cite{Bautista}, our direct calculation yields the specific combination (\ref{bdy length timelike}), which is not a matter of choice.}

{Our expression (\ref{bdy length timelike}) tends to (\ref{bdy_length_lambda=0}) in the limit $\Lambda\to 0$ if $\hat{Q}>2\alpha$. Its failure to reduce to (\ref{bdy_length_lambda=0}) in the case $\hat{Q}<2\alpha$ can be tracked to the fact that in the fixed boundary cosmological constant ensemble the limit has to be taken as discussed in (\ref{alternative limit}). This is also why we do not find the behavior proposed in \cite{Anninos} for the fixed boundary length 1-point function in the case $\hat{Q}<2\alpha$. An option to try to reproduce the behavior predicted in \cite{Anninos} would be to take $\Lambda_B'$ as the integration variable in (\ref{antiLaplace}) instead of $\Lambda_B$ in the case $\hat{Q}<2\alpha$, since that is the variable that we need to keep fixed in the limit $\Lambda\to 0$. However, we do not see any physical reason to justify so, besides that it might give the behavior predicted in \cite{Anninos}. We believe that this should be further investigated.}
\begin{figure}[h]
    \begin{center}
        \includegraphics[width=14.0
        cm]{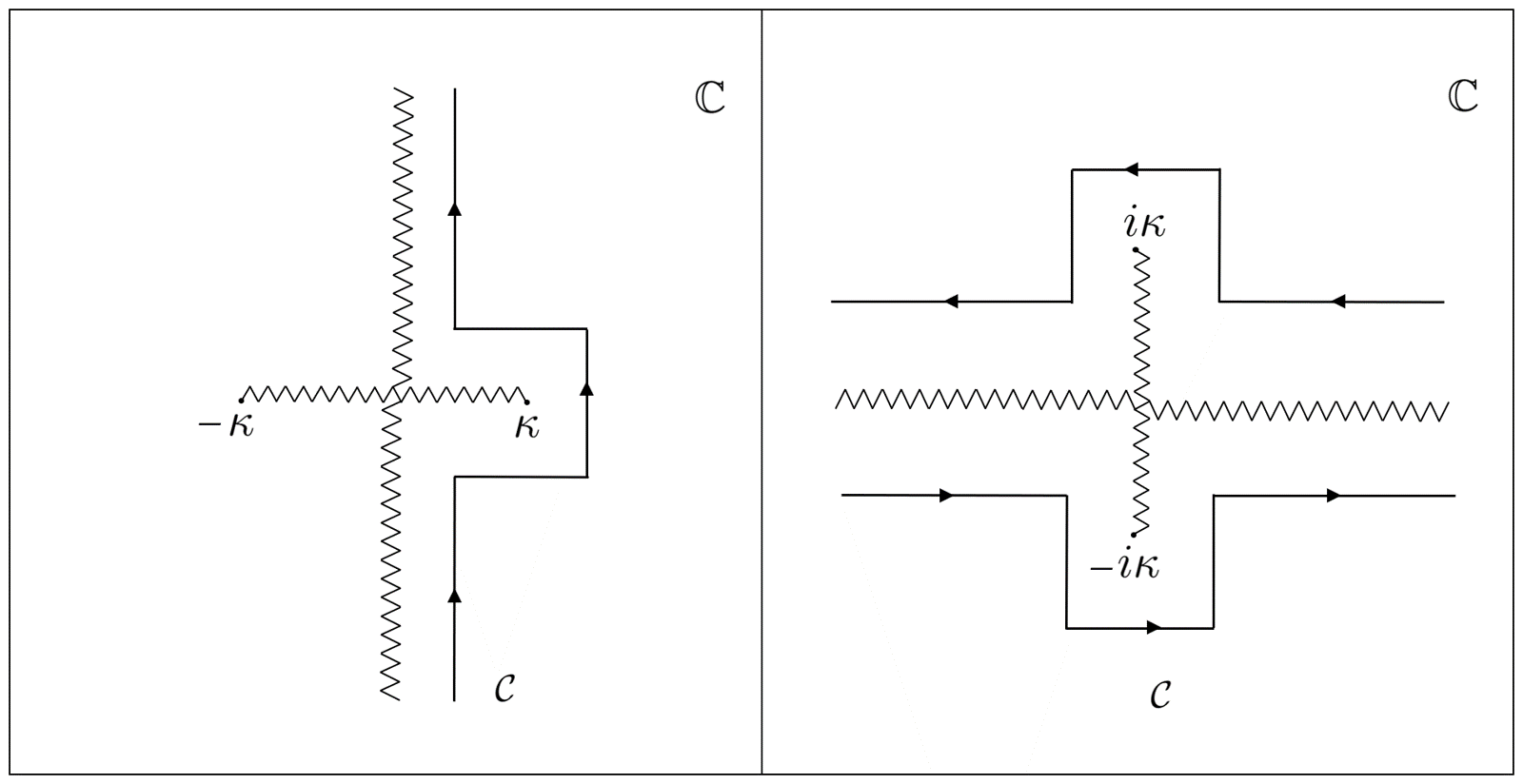}
        \caption{Scheme of how the integration contour for the integrals in $\Lambda_B\in \mathbb{C}$ involved changes when going from the spacelike (left) to the timelike (right) fixed-length 1-point function. Here, $\kappa=\sqrt{{\Lambda }/{\sin(\pi \beta^2)}}$}
        \label{Figure3}
    \end{center}
\end{figure}

\subsection{Self-duality}

We can also analyze the properties of expression (\ref{resulttimelike}) under Liouville self-duality $\beta \to -1/\beta $. To do this one defines the dual cosmological constant $\tilde \Lambda$ as follows\footnote{Cf. Eq. (3.16) in reference \cite{ZZ}. The extra phase in the timelike case is needed to ensure (\ref{DOZZt}) is self-dual.}
\begin{align}
    -\pi \tilde{\Lambda}\,\frac{\Gamma(-1/\beta^2)}{\Gamma(1+1/\beta^2)}=\left(-\pi \Lambda \,\frac{\Gamma(-\beta^2)}{\Gamma(1+\beta^2)}\,\right)^{-1/\beta^2}e^{-i\pi (1+{1}/{\beta^2})}, \label{Abajo}
\end{align}
 and the boundary dual cosmological constant $\tilde\Lambda_B$ in such a way that $\hat\gamma $ remains unchanged under $\beta \to -1/\beta$. That is,
 \begin{align}
    \text{cosh}^2(\pi \hat{\gamma}/\beta)=-\frac{\tilde{\Lambda}_B^2}{\tilde{\Lambda}}\sin(\pi/\beta^2).
    \label{dual boundary}
\end{align}
 To check that the expression (\ref{resulttimelikes}) is invariant in the case in which $(\hat{Q}-2\alpha)/\beta$ and $(\hat{Q}-2\alpha)\beta$ are integers, one has to observe that the following equation holds
\begin{equation}
{4\pi \beta }\, \frac{\Gamma(2\alpha/\beta +1/\beta ^2)}{\Gamma(2+2\alpha \beta -\beta^2 )} \, = \, \frac{4\pi }{\beta }\,\frac{\Gamma (-2\alpha \beta+\beta^2)}{\Gamma (2-2\alpha/\beta-1/\beta^2)}\, (-1)^{n+1}
\end{equation}
with $n=2\hat s(1+\beta^2)-\beta^2$. In the case where $(\hat{Q}-2\alpha)/\beta$ and $(\hat{Q}-2\alpha)\beta$ are not integers, this can also be done, although the relation is more subtle. In that case, one finds the following identity
\begin{align}
    \left(e^{-2\pi i\frac{\hat{Q}-2\alpha}{\beta}}-1\right)\hat{\text{U}}_{-{1}/{\beta}}(\alpha)=\left(e^{2\pi i\beta (\hat{Q}-2\alpha)}-1\right)\hat{\text{U}}_{\beta}(\alpha).\label{relationale}
\end{align}
To arrive at this relation one also writes $\hat{\text{U}}_{-{1}/{\beta}}(\alpha)$ in terms of the dual cosmological constants defined as in (\ref{Abajo}) and (\ref{dual boundary}}).

Therefore, we can define the normalized 1-point function as follows
\begin{equation}
\hat{\text{U}}_{\beta}(\alpha)\, \to \,\hat{\text{U}}_{\beta}'(\alpha)\,:=\,\Big(e^{2\pi i (\hat Q -2\alpha)\beta}-1\Big)\,\hat{\text{U}}_{\beta}(\alpha)\label{norma}
\end{equation}
such that it is self-dual. This normalization factor is similar to the one that appears in \cite{Santachiara}, where the case $\Lambda=0$ has been studied\footnote{See the discussion after Eq. (41) on \cite{Santachiara}.}. As we will see below, the normalized function $\hat{\text{U}}_{\beta}'(\alpha)$ satisfies both the shift equation and its dual analog. Besides, it is not difficult to see that this also satisfies the correct reflection property under $\alpha \to \hat Q-\alpha$, where now the coefficient $\text{D}(\alpha)$ has to be defined from the correlation functions of the normalized operators. 

To make the function manifestly invariant under $\beta \to -1/\beta $ one can interpret (\ref{relationale}) as suggesting the alternative normalization
\begin{equation}
\hat{\text{U}}_{\beta}(\alpha)\, \to \,\hat{\text{U}}_{\beta}''(\alpha)\,:=\,\Big(e^{\frac{-2\pi i (\hat Q -2\alpha)}{\beta }}-1\Big)^{-1}\,\hat{\text{U}}_{\beta}(\alpha)\label{norma2}
\end{equation}
which, as it is easy to prove by using properties (\ref{properties}), would result in the following expression
\begin{equation}
\hat{\text{U}}_{\beta}''(\alpha )\, =\, \frac{2 i }{(\hat Q-2\alpha )} \,
{{ \Gamma\Big(\beta ^2-2\alpha \beta  \Big)}}{{\Gamma\Big(\frac{1}{\beta^2}+\frac{2\alpha}{\beta }\Big)}}\,\Big(  \frac{\pi \Lambda\Gamma(-\beta ^2)}{\Gamma(1+\beta ^2)}\Big)^{\frac{\hat Q -2\alpha }{2\beta  }}\cosh\Big(\pi \hat{\gamma} (\hat Q - 2\alpha )\Big)\label{resulttimelikesss}
\end{equation}
which agrees with the naive analytic extension $b\to i\beta$, $\alpha\to i\alpha $, $\gamma \to i\gamma$ of expression (\ref{TheU}). {This might lead one to think that the 1-point function of the disk in the timelike theory is merely the naive analytic continuation of its spacelike analogue; however, this is not the case. Here, we are showing that the analytic continuation of the spacelike 1-point function rather corresponds to the timelike 1-point function of the re-normalized operator ${V}''_{\alpha}(z)=\frac{i\, e^{i\pi s }}{2\sin(\pi s)}e^{2\alpha\phi(z)}
$ with $s=(\hat{Q}-2\alpha )/{\beta}$, which is consistent with (\ref{resemblance}) and compatible with Eq. (11) in Ref. \cite{Santachiara}.}

\subsection{Bootstrap shift-equations}

In this section, we will show that our proposal (\ref{resulttimelikes}) for the timelike 1-point function on the disk satisfies the bootstrap shift-equations. First, let us review how it works in the spacelike case. It is not difficult to prove that the 1-point function (\ref{U}) obeys the spacelike shift equation
\begin{equation}
    -\frac{2\pi \Lambda_B}{\Gamma(-b^2)}\text{U}_b(\alpha )\, =\, \frac{\Gamma(2\alpha b -b^2)}{\Gamma(2\alpha b -1-2b^2)}\, \text{U}_b(\alpha-b/2) \, - \, \frac{\pi \Lambda \Gamma(2\alpha b-1-b^2)\Gamma(1+b^2)}{\Gamma(2\alpha b )\Gamma(-b^2)}\, \text{U}_b(\alpha +b/2)\nonumber
\end{equation}
In the same way, our formula (\ref{resulttimelikes}), which yields $\hat{\text{U}}_{\beta}(\alpha)$, obeys the analytic continuation of the above equation, defined by performing to $\alpha \to i\alpha $, $b\to i\beta$, namely 
\begin{equation}
    -\frac{2\pi \Lambda_B}{\Gamma(\beta^2)}\hat{\text{U}}_{\beta}(\alpha )
    \, =\, \frac{\Gamma(\beta^2 - 2\alpha \beta)}{\Gamma(2\beta^2-1-2\alpha \beta )}\, \hat{\text{U}}_{\beta}(\alpha-\beta/2) 
    \, - \, \frac{\pi \Lambda \Gamma(\beta^2-1-2\alpha \beta)\Gamma(1-\beta^2)}{\Gamma(-2\alpha \beta )\Gamma(\beta^2)}\, \hat{\text{U}}_{\beta}(\alpha +\beta/2)
    \label{shift eq timelike}
\end{equation}
More interestingly, the normalized function $\hat{\text{U}}'_{\beta}(\alpha)$, defined as in (\ref{norma}), satisfies the corresponding shift equation with the new normalization factors\footnote{The shift-equations are sensitive to the normalization of the operators, so in order to work with operators with a different normalization, one has to make the corresponding changes to the shift-equations.}
\begin{eqnarray}
    -\frac{2\pi \Lambda_B}{\Gamma(\beta^2)}\hat{\text{U}}'_{\beta}(\alpha )
    \, &=&\,-\,e^{-i\pi \beta^2}\, \frac{\Gamma(2+2\alpha\beta -2\beta^2)}{\Gamma(1-\beta^2+2\alpha\beta )}\, \hat{\text{U}}'_{\beta}(\alpha-\beta/2) 
    \,  \\ 
    &&\, + \,e^{i\pi\beta^2}\, \frac{\pi \Lambda \Gamma(1+2\alpha \beta )\Gamma(1-\beta^2)}{\Gamma(2+2\alpha\beta -\beta^2)\Gamma(\beta^2)}\, \hat{\text{U}}'_{\beta}(\alpha +\beta/2)\nonumber
\end{eqnarray}
together with its dual counterpart 
\begin{eqnarray}
    -\frac{2\pi \tilde\Lambda_B}{\Gamma(1/\beta^2)}\hat{\text{U}}'_{\beta}(\alpha )
    \, &=&\,-\,e^{-i\pi /\beta^2}\, \frac{\Gamma(2-2\alpha/\beta -2/\beta^2)}{\Gamma(1-1/\beta^2-2\alpha/\beta )}\, \hat{\text{U}}'_{\beta}(\alpha+1/2\beta ) 
    \, 
     \\ 
    &&\,+ \,e^{i\pi/\beta^2}\, \frac{\pi \tilde\Lambda \Gamma(1-2\alpha/ \beta )\Gamma(1-1/\beta^2)}{\Gamma(2-2\alpha/\beta -1/\beta^2)\Gamma(1/\beta^2)}\, \hat{\text{U}}'_{\beta}(\alpha -1/2\beta )\nonumber
\end{eqnarray}
for (\ref{Abajo}) and (\ref{dual boundary}). Notice that because the normalization factor (\ref{norma}) remains invariant if we change $\alpha \to \alpha \pm 1/2\beta$, the above equation also holds for the 1-point function without the normalization (\ref{resulttimelikes}).

Equivalently, the function $\hat{\text{U}}_{\beta}(\alpha)$, in addition to satisfying (\ref{shift eq timelike}), also satisfies
\begin{align}
    -\frac{2\pi \tilde{\Lambda}_B}{\Gamma(1/\beta^2)}\frac{\hat{\text{U}}_{\beta}(\alpha )}{\sin(\pi(\hat{Q}-2\alpha)/\beta)}
    \, =&\, e^{-i\pi/\beta^2}\frac{\Gamma(1/\beta^2 + 2\alpha/ \beta)}{\Gamma(2/\beta^2-1+2\alpha/ \beta )}\, \frac{\hat{\text{U}}_{\beta}(\alpha+1/2\beta)}{\sin(\pi(\hat{Q}-2\alpha-1/\beta)/\beta)} 
    \, \label{shift eq timelike_improved_dual}\\-& \, e^{i\pi/\beta^2}\frac{\pi \tilde{\Lambda} \Gamma(1/\beta^2-1+2\alpha/ \beta)\Gamma(1-1/\beta^2)}{\Gamma(2\alpha/ \beta )\Gamma(1/\beta^2)}\, \frac{\hat{\text{U}}_{\beta}(\alpha -1/2\beta )}{\sin(\pi(\hat{Q}-2\alpha+1/\beta)/\beta)}.
    \nonumber 
\end{align}
Notice the extra factors $\sin(\pi 2\hat{s}\pm\pi/\beta^2)$ and the phases that appear in this expression and compare them with the expression obtained by just taking $\beta\to-1/\beta$, $\Lambda\to\tilde{\Lambda}$ and $\Lambda_B\to\tilde{\Lambda}_B$ in (\ref{shift eq timelike}). This is also consistent with the result in \cite{Santachiara}, where the same trigonometric functions appear\footnote{See the shift-equations below Eq. (41) in \cite{Santachiara}.} when $\Lambda =0$.

This explains why our result for the timelike disk 1-point function (\ref{resulttimelikes}) differs from the one in \cite{Bautista}; in the latter work the authors found a 1-point function that satisfies the analytic extension of both the spacelike shift-equation and its dual. Both here and in \cite{Santachiara} it has been shown that the solution stemming from the path integral formulation satisfies the analytic continuation of the spacelike shift-equation, but not exactly the analytical continuation of the spacelike dual shift-equation, but a similar equation with extra trigonometric factors. It would be interesting to understand from the bootstrap point of view why the dual shift-equation picks up these extra factors. Notice that one ingredient needed to derive the dual shift equation is the operator product expansion between the bulk operator $V_{-\frac{1}{2\beta}}$ and the boundary identity. The corresponding bulk-boundary 2-point function in the timelike case is yet to be computed.

\section{Summary}

In this paper, we have derived an explicit formula for the disk 1-point function for the timelike Liouville theory. In order to do that, we used the Coulomb gas formalism and analytically extended the expression in the number of screening operators. Our result reads
\begin{equation}
 \big\langle \,V_{\alpha}(z, \bar z)\,\big\rangle \,= \,\big|z-\bar z \big|^{2\alpha (\hat{Q}-\alpha )}\,\hat{\text{U}}_{\beta}(\alpha )
\end{equation}
with $V_{\alpha}(z, \bar z)=e^{2\alpha \phi(z, \bar z)}$, where
%\begin{equation}
%\hat{\text{U}}_{\beta}(\alpha )\, =\, \frac{4\pi }{\beta} \,e^{- \frac{i\pi(\hat Q-2\alpha )}{2\beta}}\,
%\frac{{ \Gamma\Big(\beta ^2-2\alpha \beta  \Big)}}{{\Gamma\Big(2-\frac{2\alpha}{\beta } - \frac{1}{\beta ^2}\Big)}}\Big(  -\frac{\pi \Lambda\Gamma(-\beta ^2)}{\Gamma(1+\beta ^2)}\Big)^{\frac{\hat Q -2\alpha }{2\beta  }}\cosh\Big(\pi \hat{\gamma} (\hat Q - 2\alpha )\Big)\label{resulttimelikess}
%\end{equation}
\begin{equation}
\hat{\text{U}}_{\beta}(\alpha )\, =\, \frac{4\pi }{\beta}e^{-i\pi\frac{\hat{Q}-2\alpha}{2\beta}} \,
\frac{{ \Gamma\Big(\beta ^2-2\alpha \beta  \Big)}}{{\Gamma\Big(2-\frac{2\alpha}{\beta } - \frac{1}{\beta ^2}\Big)}}\, \Bigg(-  \frac{\pi \Lambda\Gamma(-\beta ^2)}{\Gamma(1+\beta ^2)}\Bigg)^{\frac{\hat Q -2\alpha }{2\beta  }}\cosh\Big(\pi \hat{\gamma} (\hat Q - 2\alpha )\Big)\label{resulttimelikess}
\end{equation}
with $\hat{Q}=\beta-1/\beta $ and $\hat \gamma $ being defined as $\cosh^2(\pi \beta  \hat{\gamma} )= -({\Lambda_B^2}/{\Lambda })\sin(\pi \beta ^2)$. We have shown that this formula satisfies the following properties:
\begin{itemize}
\item In the limit $\Lambda \to 0$, it reproduces the result obtained in \cite{Santachiara}.
    \item {It connects to the path integral formulation of the field theory.}
    \item It satisfies the right reflection relation under $\alpha \to \hat{Q}-\alpha$, as the timelike DOZZ formula. 
    \item It obeys the bootstrap shift-equations that appear in the 2-dimensional bootstrap.
    \item It has nice properties under self-duality $\beta \to -1/\beta$.
\end{itemize}
It is also important to remark that
\begin{itemize}
    \item Our result differs from the expression proposed in \cite{Bautista}, especially in the dependence of $\Lambda_B$.
\end{itemize}

The motivation we had to look at the timelike 1-point function was twofold. On the one hand, this is important to further investigate the timelike version of Liouville theory, which is much less understood than its spacelike counterpart. On the other hand, this is important to explore the applications of this theory as a toy model to study quantum cosmology in de Sitter space. 

As a future direction, we would like to extend the analysis of \cite{Santachiara} to the case $\Lambda \neq 0$. The Coulomb gas formalism has a clear limitation, as it involves calculating for configurations corresponding to integer values of the number of screening operators, which requires an analytical extension of the resulting expressions. In contrast, the formalism of \cite{Santachiara} would allow for a calculation for generic values of the momentum $\alpha$ without the need for analytical extension. However, generalizing such a method to non-vanishing $\Lambda $ is more complicated. This is a matter for future work.

{It would also be interesting to fully understand the relation between this quantity and quantum cosmology, and to reconcile our results to the arguments of \cite{Anninos}.} In other words, it would be interesting to identify what is the contour prescription that leads to the wave function of the universe proposed in \cite{Anninos}, and what is the physical justification for such a choice from the 2D CFT point of view.

%\[\]
\subsection*{Acknowledgments}
The authors thank Raoul Santachiara, Yifan Wang, Edward Witten and Themistocles Zikopoulos for useful discussions. Research at Perimeter Institute is supported in part by the Government of Canada through the Department of Innovation, Science and Economic Development and by the Province of Ontario through the Ministry of Colleges, Universities, Research Excellence and Security.

\end{document}